\documentclass[aps,prd,twocolumn,amsfonts,amssymb,amsmath,showpacs,letterpaper,superscriptaddress,nofootinbib,altaffilletter,fontenc]{revtex4-2}

\usepackage[citecolor=blue,linkcolor=blue,colorlinks=true,urlcolor=blue]{hyperref}

\usepackage[utf8]{inputenc}
\usepackage[T1]{fontenc}
\usepackage{graphicx}
\usepackage{amssymb}
\usepackage{amsmath}
\usepackage{aas_macros}
\usepackage{rotating}
\usepackage{color}
\usepackage{epsf}
\usepackage{fancyhdr}
\usepackage{ifthen}
\usepackage{slashed}
\usepackage{xspace}
\usepackage{tabularx}
\usepackage{multirow}
\usepackage{array}
\usepackage{makecell}
\usepackage{caption}
\usepackage{subcaption}
\usepackage{float}

\usepackage[misc]{ifsym}
\usepackage{orcidlink}

\newcolumntype{M}[1]{>{\centering\arraybackslash}m{#1}}

\begin{document}

\title{Search for gravitational-wave bursts in the third Advanced LIGO-Virgo run with coherent WaveBurst enhanced by Machine Learning}




\author{Marek J. Szczepa\'nczyk \orcidlink{0000-0002-6167-6149}}
\affiliation{Department of Physics, University of~Florida, PO Box 118440, Gainesville, FL 32611-8440, USA}
\author{Francesco Salemi \orcidlink{0000-0002-9511-3846}}
\email[E-mail: ]{francesco.salemi@unitn.it}
\author{Sophie Bini \orcidlink{0000-0002-0267-3562}}
\affiliation{Universit\`a di Trento, Dipartimento di Fisica, I-38123 Povo, Trento, Italy}
\affiliation{INFN, Trento Institute for Fundamental Physics and Applications, I-38123 Povo, Trento, Italy}
\author{Tanmaya Mishra \orcidlink{0000-0002-7881-1677}}
\affiliation{Department of Physics, University of~Florida, PO Box 118440, Gainesville, FL 32611-8440, USA}
\author{Gabriele Vedovato \orcidlink{0000-0001-7226-1320}}
\address {Universit\`a di Padova, Dipartimento di Fisica e Astronomia, I-35131 Padova, Italy}
\author{V. Gayathri \orcidlink{0000-0002-7167-9888}}
\affiliation{Department of Physics, University of~Florida, PO Box 118440, Gainesville, FL 32611-8440, USA}
\author{Imre Bartos \orcidlink{0000-0001-5607-3637}}
\affiliation{Department of Physics, University of~Florida, PO Box 118440, Gainesville, FL 32611-8440, USA}
\author{Shubhagata Bhaumik \orcidlink{0000-0001-8492-2202}}
\affiliation{Department of Physics, University of~Florida, PO Box 118440, Gainesville, FL 32611-8440, USA}
\author{Marco Drago \orcidlink{0000-0002-3738-2431}}
\affiliation{Universit\`a di Roma  La Sapienza, I-00185 Roma, Italy}
\affiliation{INFN, Sezione di Roma, I-00185 Roma, Italy}
\author{Odysse Halim \orcidlink{0000-0003-1326-5481}}
\affiliation{Dipartimento di Fisica, Universit\`a di Trieste, I-34127 Trieste, Italy}
\affiliation{INFN, Sezione di Trieste, I-34127 Trieste, Italy}
\author{Claudia Lazzaro \orcidlink{0000-0001-5993-3372}}
\affiliation{Universit\`a di Padova, Dipartimento di Fisica e Astronomia, I-35131 Padova, Italy }
\affiliation{INFN, Sezione di Padova, I-35131 Padova, Italy }
\author{Andrea Miani \orcidlink{0000-0001-7737-3129}}
\affiliation{Universit\`a di Trento, Dipartimento di Fisica, I-38123 Povo, Trento, Italy}
\affiliation{INFN, Trento Institute for Fundamental Physics and Applications, I-38123 Povo, Trento, Italy}
\author{Edoardo Milotti \orcidlink{0000-0001-7348-9765}}
\affiliation{Dipartimento di Fisica, Universit\`a di Trieste, I-34127 Trieste, Italy}
\affiliation{INFN, Sezione di Trieste, I-34127 Trieste, Italy}
\author{Giovanni A. Prodi \orcidlink{0000-0001-5256-915X}}
\affiliation{Universit\`a di Trento, Dipartimento di Matematica, I-38123 Povo, Trento, Italy}
\affiliation{INFN, Trento Institute for Fundamental Physics and Applications, I-38123 Povo, Trento, Italy}
\author{Shubhanshu Tiwari \orcidlink{0000-0003-1611-6625}}
\affiliation{Physik-Institut, University of Zurich, Winterthurerstrasse 190, 8057 Zurich, Switzerland}
\author{Sergey Klimenko \orcidlink{0000-0003-0710-4331}}
\affiliation{Department of Physics, University of~Florida, PO Box 118440, Gainesville, FL 32611-8440, USA}

\begin{abstract}

This paper presents a search for generic short-duration gravitational-wave (GW) transients (or GW bursts) in the data from the third observing run of Advanced LIGO and Advanced Virgo. 
We use coherent WaveBurst (cWB) pipeline enhanced with a decision-tree classification algorithm for more efficient separation of GW signals from noise transients. The machine-learning (ML) algorithm is trained on a representative set of noise events and  a set of simulated stochastic signals that are not correlated with any known signal model. This training procedure preserves the model-independent nature of the search. 
 We demonstrate that the ML-enhanced cWB pipeline can detect GW signals at a larger distance than previous model-independent searches. 
The sensitivity improvements are achieved across the broad spectrum of simulated signals, with the goal of testing the robustness of this model-agnostic search. 
 At a false-alarm rate of one event per century, 
the detectable signal amplitudes are reduced up to almost an order of magnitude, 
most notably for the single-cycle signal morphologies.
This ML-enhanced pipeline also improves the detection efficiency of compact binary mergers in a wide range of masses, from stellar mass to intermediate-mass black holes, both with circular and elliptical orbits.
After excluding previously detected compact binaries, no new gravitational-wave signals are observed for the two-fold Hanford-Livingston and the three-fold Hanford-Livingston-Virgo detector networks.
 With the improved sensitivity of the all-sky search, we obtain the most stringent constraints on the isotropic emission of gravitational-wave energy from short-duration burst sources.  

\end{abstract}

\date[\relax]{Dated: \today }

\maketitle


\section{Introduction}


The first detection of a gravitational wave (GW) signal, GW150914 ~\cite{GW150914}, in 2015, during the earliest observing run of the Advanced LIGO detectors~\cite{TheLIGOScientific:2014jea}, opened a new observation window to study the Universe. In the following runs, the Advanced LIGO and Advanced Virgo~\cite{TheVirgo:2014hva} detectors have recorded around a hundred GW signals~\cite{GWTC1, GWTC2, GWTC2.1, GWTC3}. All these detected GW sources are interpreted as compact binary coalescences (CBC), with the majority being binary black hole (BBH) mergers. However, gravitational waves from other, less understood GW sources, still remain elusive~\cite{O3allskyshort,KAGRA:2021bhs}. 

Several astrophysical sources are predicted to produce short bursts of GW radiation with a duration of up to a few seconds. They could be roughly divided into two categories. The first category includes late inspiral and merger waves from CBCs comprised of black holes and/or neutron stars. Types of CBC signals include binary systems with circular orbits ~\cite{2004cbhg.symp...37V,2004IJMPD..13....1M,Ebisuzaki:2001qm,2017mbhe.confE..51K,Inayoshi:2019fun}, eccentric orbits~\cite{2017MNRAS.468.5020I,2016A&A...588A..50M,2016MNRAS.460.3545D,2000ApJ...528L..17P,2017ApJ...835..165B,GayathriNatAstro,SamsingNat}, head-on collisions~\cite{Bustillo:2020ukp,Ebersold:2022zvz}, extreme mass-ratio binaries~\cite{East:2014nfa}, primordial black holes~\cite{Sasaki:2018dmp,Green:2020jor,Franciolini:2021nvv,Ebersold:2020zah}, and hyperbolic encounters~\cite{DeVittori:2012da,Cho:2018upo,Codazzo:2022aqj}. Gravitational waves may also be generated in the post-merger phase of the binary neutron star systems~\cite{Abbott:2017dke,Benhar:2004xg}. The second category includes all other anticipated burst progenitors, such as star explosions, neutron star glitches, or unknown sources. Prime examples of the star explosions are core-collapse supernovae (CCSNe)~\cite{Mezzacappa:2020lsn,Powell2020,Radice2019,Andresen2019,OConnor2018,Kuroda2017,Obergaulinger:2020cqq,Szczepanczyk:2021bka}, hypernovae~\cite{2009PhRvD..80l3008S}, superluminous supernovae~\cite{Cheng:2018awt}, supernovae type Ia~\cite{Stephan:2019fhf},  and pulsation pair-instability supernovae~\cite{Fryer:2000my,Powell:2021qib,Rahman:2021dvs}. Also, gravitational waves may be produced during a black-hole formation from a collapsing star~\cite{Cerda-Duran2013,Kuroda2018,Pan2020}, quantum chromodynamics phase transition in a core-collapse supernova~\cite{Kuroda:2021eiv}, or in a collapsar~\cite{Wei:2019ljb,Cerda-Duran:2013swa}.  The resulting gravitational waves from these source types are typically predicted to be stochastic.
Neutron stars or pulsar glitches are expected to generate ringdown GW signals~\cite{mech_NS,superfluid,sf2,Lopez:2022yph}. Other potential burst sources are the soft gamma repeaters~\cite{LIGOScientific:2008nqs} and cosmic strings (CSs)~\cite{Kibble:1976sj,Damour:2004kw}.


For most of the GW sources mentioned above, the simulations of accurate and computationally efficient GW waveforms (templates) are not readily available, which limits the use of the matched filter methods. 
This is the reason why the burst algorithms~\cite{PhysRevD.93.042004,CWBwavelet:2004km,CWBlikenet:2005kmrm,SoftwareX, zenodo,Lynch:2015yin,Cornish2:2014kda}, designed to detect a wide range of GW sources without templates, are actively used for GW searches in data collected by the LIGO\cite{TheLIGOScientific:2014jea}, Virgo~\cite{TheVirgo:2014hva}, KAGRA~\cite{KAGRA:2013pob} and GEO600~\cite{Grote:2010zz} detectors. 

Coherent WaveBurst (cWB)~\cite{PhysRevD.93.042004} has been used to search for burst signals~\cite{Abadie:2012rq,Abadie:2010mt,Abbott:2009zi,Abbott:2005at,Abbott:2005rt,Abbott:2005fb,Abbott:2004rt} and BBH signals~\cite{GWTC1,GWTC2,GWTC3,Abbott:2017iws,Abbott:2019ovz,Virgo:2012aa,aasi:2014iwa,Salemi:2019owp}. It has made a major contributions in the discovery of the first GW signal GW150914~\cite{GW150914,LIGOScientific:2016fbo} and the first intermediate-mass black hole merger GW190521~\cite{Abbott:2020tfl,GW190521_cwb}. Also, cWB contributed to the Gravitational Wave Transient Catalogs (GWTCs)~\cite{GWTC1, GWTC2,GWTC3}, to population studies of IMBH sources~\cite{POP_GWTC2,OBrien_2021}, and the search for specific CBC properties, such as higher-order multipoles~\cite{GW190814,vedovato2022minimally}. 

By using a few assumptions about GW signals, 
cWB identifies the excess power triggers 
in the data from multiple detectors. 
 A reduction of the false-alarm rate is accomplished with a post-production 
 veto procedure making use of summary statistics calculated for each trigger.
This paper presents a generic burst search using cWB enhanced with machine learning (ML) to improve the separation of GW signals from the noise transients. 
Similar to the standard veto procedure, supervised classification algorithms can  achieve an efficient reduction of the cWB false-alarm rate. 
In~\cite{Vinciguerra_2017} a neural network method analyzing the time-frequency patterns of reconstructed cWB triggers was suggested to improve the detection of BBH signals. In~\cite{Cavagli__2020} a machine-learning method was used to improve the identification of CCSNe. In both cases, a strong model dependence was introduced at the cWB post-production stage, limiting the pipeline detection to a specific source used for the ML training. A more practical weakly-modeled classification based on the XGBoost decision trees was introduced in~\cite{XGBcwb, O3XGB} to enhance the cWB detection of the BBH signals. Unlike template-based searches that find events based on a specific signal model, the XGBoost classification is designed to penalize events inconsistent with generic signal features.
More recently, a different signal-noise classification based on the Gaussian Mixture Model (GMM) methodology was developed for generic burst searches with cWB~\cite{gmmcwb2022}.
In this approach, the GMM likelihood model is tuned on a separated subset of the simulated events used to benchmark the performances, sparsely sampling the ranges of burst frequency and duration.
Instead, in this paper, we use the XGBoost decision-tree algorithm, which is trained on a representative set of noise events and a set of simulated signals with stochastic waveforms that are not correlated with any known signal model. The training set is densely covering the frequency band and the range of burst duration selected for the analysis. We demonstrate the sensitivity improvement across a broad spectrum of simulated signals not used for the XGBoost training.
With this more sensitive algorithm, we re-analyze the data from the third observing run (O3) of the LIGO and Virgo detectors, targeting bursts in the frequency band up to 1\ kHz.

The paper is structured as follows. Section ~\ref{sec:method} describes the ML-enhanced cWB pipeline, the data analyzed, and simulated signals used for training the ML algorithm. Section~\ref{sec:models} describes signal models used for testing the search sensitivity. Section~\ref{sec:results} provides results of the ML-enhanced cWB search compared with the LIGO-Virgo-KAGRA (LVK) all-sky search for short duration bursts~\cite{O3allskyshort}. 

\section{Method}
\label{sec:method}

\subsection{Coherent WaveBurst pipeline enhanced with machine learning}
\label{sec:ML}

Coherent WaveBurst (cWB) is a burst search pipeline 
that can detect GW signals from astrophysical sources without templates~\cite{PhysRevD.93.042004,CWBwavelet:2004km,CWBlikenet:2005kmrm,SoftwareX, zenodo}.
The cWB algorithm identifies excess-power triggers in the time-frequency data
obtained with the multi-resolution Wilson-Daubechies-Meyer (WDM) wavelet transform~\cite{Necula:2012zz} of the strain data from multiple detectors. 
By using the constrained likelihood method~\cite{PhysRevD.93.042004}, it reconstructs the source location in the sky and the signal waveform: 
incoming gravitational waves should produce coherent responses over the detector network, while in general, the noise events are not correlated.
The rate of triggers produced by the pipeline is controlled by model-independent veto thresholds on the excess-power statistics and the coherent amplitude $\eta_\mathrm{0}$ that characterizes the coherence of a cWB trigger across the detector network. The veto thresholds are set to be sufficiently low to avoid costly re-runs of the pipeline trigger production. 
Further reduction of the cWB false-alarm rate is achieved with a more stringent veto procedure by applying thresholds 
on the summary statistics calculated for each trigger, including the $\eta_\mathrm{0}$, the cross-correlation coefficient between the detectors, the reconstructed signal-to-noise ratio in each detector, the number of WDM resolutions used for the reconstruction and the waveform shape parameters (see Appendix \ref{sec:xgb}for more details).  

Although this veto procedure generally works quite efficiently, designing vetoes in the multi-dimensional space of summary statistics is challenging and requires re-tuning the veto thresholds for each detector network configuration and observing run. In addition, due to distinctive noise sources, cWB triggers are typically split into different categories (so-called "search bins") and analyzed separately. 
To solve these issues, a boosted decision-tree algorithm, eXtreme-Gradient Boost (XGBoost)~\cite{XGBoost}, was adopted and implemented within the cWB framework to automate the signal-noise classification of cWB triggers~\cite{XGBcwb}. 
Two types of input data are used: signal events from simulations and noise events from background estimations. For each event, a selected subset of cWB summary statistics (see Appendix ~\ref{sec:xgb} for more details) is used by XGBoost as input features to train a signal-noise model.
This ML method simplifies the analysis approach 
by jointly analyzing all candidate events within a single search bin. 
 As described in Refs.~\cite{XGBcwb,O3XGB}, the detection statistic for the ML-enhanced cWB algorithm is defined by:
\begin{equation}\label{eq:x}
    \eta_\mathrm{r} = \eta_\mathrm{0}\cdot W_{\mathrm{XGB}}, 
\end{equation}
where $\eta_\mathrm{0}$ is the cWB ranking statistic and $W_{\mathrm{XGB}}$ is the XGBoost penalty factor ranging between 0 (noise) and 1 (signal).

This study extends the ML-enhanced cWB method~\cite{XGBcwb,O3XGB} to generic burst searches in the frequency band [16, 1024]\,Hz. In order to preserve the model-independent nature of cWB and be sensitive to generic GW signals, when training the XGBoost, we select as input features the summary statistics that do not depend on the waveform morphology (see Appendix~\ref{sec:xgb} for details). Moreover, we do not train XGBoost on anticipated GW signals that follow specific astrophysical distributions. Instead, we use a stochastic set of band-limited White-Noise-Burst (WNB) signals: i.e., white noise contained within an ad-hoc time-frequency range. The WNB signals populate a wide range of randomly chosen signal durations and frequency bands. To that end, we use two sets of WNB for training: (a) WNBs uniformly distributed in central frequency in the range [24,996] Hz, bandwidth [10,300]\,Hz, and duration is logarithmically distributed between 0.1\,ms and 1\,ms; (b) WNBs logarithmically distributed in central frequency [24,450]\,Hz, bandwidth [10,250]\,Hz and duration [1,50]\,ms. 
The WNB signals are explicitly chosen to span a space of two parameters ($Q_\mathrm{a}$, $Q_\mathrm{p}$): 
$Q_\mathrm{a}$ is an estimator of how much energy occurs outside the largest oscillation of the event's waveform
and $Q_\mathrm{p}$ is a parameter linked to the event number of cycles~\cite{XGBcwb,O3XGB}. Within this simplified 2-D parameters space, low-$Q$ noise transients can be better identified by the ML classifier and, therefore, penalized. Furthermore, this approach efficiently removes one of the most challenging noise sources - blip glitches~\cite{Cabero:2019orq,O3allskyshort}. The details of the changes introduced to the ML method for improving the sensitivity toward generic burst sources are described in Appendix~\ref{sec:xgb}.

\subsection{Data}

The third observing run of the Advanced LIGO, Hanford (H) and Livingston (L), and Advanced Virgo (V) detectors consists of two epochs separated by a commissioning break: O3a (from April 1, 2019, to October 1, 2019) and O3b (from November 1, 2019, to March 27, 2020). The two epochs have significantly different rates of short-duration glitches. 

In this paper, we analyze HL and HLV detector networks. The HV and LV networks are omitted due to their lower sensitivities and shorter additional observing time with respect to the HL network. The three-fold network generally allows a more accurate characterization of the GW signal than a two-detector network. 
Previous burst searches on O3 data, in particular the LVK one~\cite{O3allskyshort}, did not report analysis of the HLV detector network because its detection performances were found to be not as good as the ones achieved using the HL detector network. This is due to concurrent causes, including differences in spectral and directional sensitivities between the two LIGO detectors and Virgo.
After removing periods affected by poor environmental conditions and detector hardware anomalies, the analyzed duration of coincident data between LIGO detectors is 206.57 days (104.94 and 101.63 days for O3a and O3b, respectively). In comparison, it is 143.3 days (75.19 and 68.11 days) for HLV. We calculated the background data set of our search by using the time-shift analysis. We accumulated 980.7 and 1096.0~years of data for the HL network for O3a and O3b, respectively, while it was 572.9 and 395.8~years for the HLV network. Around 50\% of the background data is used for training the XGBoost model.

\subsection{Metric statistics for comparing performances}
\label{sec:metric}

The sensitivity of searches for GW bursts is usually measured in terms of detection efficiency as a function of root-squared-sum (rss) strain amplitude of the signal, $h_{\mathrm{rss}}$ \footnote{More precisely, for isotropic emission $h_{\mathrm{rss}}$ indicates signal strength at Earth, while for an-isotropic emissions we define $h_{\mathrm{rss}}$ as the maximum amplitude emitted by the source, inversely scaled by the source distance, as in~\cite{O3allskyshort}}:
\begin{equation}
 h_{\mathrm{rss}} = \sqrt{\int_{-\infty}^{\infty} \left[h_+^2(t)+h^2_\times(t)\right] \, \mathrm{d}t}\,,
 \label{eq:hrss}
\end{equation}
where $h_+$ and $h_\times$ are polarization components. Detection efficiency $\epsilon (h_\mathrm{rss})$ is estimated by adding (i.e. injecting) selected GW signals in the detector noise over a wide range of amplitudes. For all-sky searches, the modeled source location and orientation in the sky are uniformly random. The resulting
$\epsilon (h_\mathrm{rss})$ is the fraction of detected events over injected ones at  $h_{\mathrm{rss}}$ amplitude with a search threshold that ensures a minimum statistical significance, such as a minimum inverse False Alarm Rate (iFAR). 

The $h_{\mathrm{rss}}$ amplitude corresponding to 50\% detection efficiency, $h_{\mathrm{rss50}}$, has been widely used as a benchmark of the typical search sensitivity. Here we report $h_{\mathrm{rss50}}$ at iFAR $\geq 100$~years as in~\cite{O3allskyshort}, i.e. for rather significant detection candidates.

Detection efficiency can also be expressed as a function of the luminosity distance $r$ by assuming a reference amplitude value of the GW emission, $h_\mathrm{rss}(r_0)$  at some nominal distance $r_0$:
\begin{equation}
    \epsilon(r) \approx \epsilon\left(h_\mathrm{rss}(r_0)\right) \times h_\mathrm{rss}(r_0) / h_\mathrm{rss} (r).
    \label{eqn:dist}
\end{equation}
This approximation is appropriate whenever the emission process of interest is close to that of a standard siren or its model sets a specific energy scale (as for CCSNe models). Eq.~\ref{eqn:dist} also describes how detection efficiency $\epsilon(r)$ scales for different assumptions on the $(r_0, h_\mathrm{rss}(r_0))$ parameters, provided that waveform morphology and spectral sensitivities of detectors are unchanged.

Detectable volume, or sensitivity volume $\mathcal{V}$, is a benchmark more directly related to the detection probability of a source population, whose spatial and GW amplitude distributions are assumed to be known. In case the source is modeled as a standard siren with a uniform spatial density and rate, the sensitive volume can be estimated by integrating the detection efficiency\footnote{This formula is valid for source distances at negligible cosmological redshift.} (see~\cite{Abadie:2012rq}):
\begin{equation}
    \mathcal{V} = 4 \pi (r_0 \ h_\mathrm{rss}(r_0))^3 \int_0^\infty \frac{d h_\mathrm{rss}}{h_\mathrm{rss}^4} \epsilon (h_\mathrm{rss})\,
\label{eq:vol}
\end{equation}
and volume $\mathcal{V}$ can be re-scaled for different  amplitude-distance relations $h_\mathrm{rss}(r_0)$. 

The following expression provides the energy radiated in GWs 
in case the emission is narrowband and isotropic:
\begin{equation}
\label{eq:eng}
  E_\mathrm{GW} = \frac{\pi^2 c^3}{G} r^2_0 f^2_0 h_\mathrm{rss}^2(r_0)
\end{equation}
where $r_0$ is the distance to the source and $f_{0}$ is the central frequency of the GW signal. This approximation is good to within a few \% for all the ad-hoc signals listed in subsection~\ref{subsec:adhoc}, apart from the broad-band Gaussian pulses. 
By using $h_{\mathrm{rss50}}$ in Eq.~\eqref{eq:eng}, one can benchmark the typical sensitivity of the search in terms of $E_\mathrm{GW}$ at a reference distance (50\% detection efficiency, for the selected iFAR threshold). Conversely, assuming a source emitting $E_\mathrm{GW}$ around a peak frequency $f_0$, Eq.~\eqref{eq:eng} can be used to estimate the product $r_0 \  h_\mathrm{rss}(r_0)$:
\begin{equation}
    r_0 \ h_\mathrm{rss}(r_0) = \frac{\sqrt{G}}{\pi \sqrt{c^3} f_0} \sqrt{E_\mathrm{GW}}.
\end{equation}
and knowledge of the $h_{\mathrm{rss50}}$ value allows estimating the typical range of the search.

\section{Signal models}
\label{sec:models}

The generic burst searches are designed to be sensitive to a wide range of GW morphologies, and the ML-enhanced cWB is tested with various possible burst signals. This Section describes ad-hoc signals and the waveforms derived from several astrophysical models (namely CCSNe, ringdowns, and cosmic strings) for a total of 53 tested signal morphologies, listed in Table~\ref{tab:hrss}. For robustness, sensitivity to different types of compact binary systems is also studied.

\subsection{Ad-hoc}
\label{subsec:adhoc}

The ad-hoc signals estimate the algorithm's sensitivity to generic GW morphologies. They include sine-Gaussians (SGs), Gaussian pulses (GAs), and WNBs. The SGs are fully characterized by the central frequency $f_0$ and quality factor $Q$ determining the signal's duration. The GAs have only one parameter, i.e., the duration of one standard deviation $\tau_\mathrm{GA}$. Finally, the WNBs are described by a lower frequency bound $f_\mathrm{low}$, a bandwidth $\Delta f$, and a duration $\tau_\mathrm{WNB}$. More details on these signal morphologies can be found in~\cite{O3allskyshort,Abadie:2012rq}.

\subsection{Astrophysical}
\label{subsec:astro}

One of the most interesting astrophysical burst sources is CCSNe. These exploding stars are very challenging to model since all forces of nature on a micro- and macro-scale impact the explosion. GWs from CCSNe are stochastic, and their typical duration is of the order of 1\,s. The typical energies range between $10^{-10} M_\odot c^2$ and $10^{-7} M_\odot c^2$ with peak frequencies at around 1\,kHz. More details about the properties of GWs from CCSNe can be found in~\cite{Szczepanczyk:2021bka}. We study waveforms from various models that explore progenitor star rotation and masses, explosion phases, energies and GW signatures. The 10 analyzed neutrino-driven explosion models are: Andresen et al. 2017 \cite{Andresen_2017} (And+17) s11, Kuroda et al. 2016 \cite{Kuroda2016} (Kur+16) TM1, M\"uller et al. 2012 \cite{Muller2012} (Mul+12) L15,  O'Connor \& Couch 2018 \cite{OConnor2018} (Oco+18) mesa20, Powell \& M\"uller 2019 \cite{Powell2019} (Pow+19) he3.5 and s18, Radice et al. 2019  \cite{Radice2019} (Rad+19) s9, s13, s25. Additionally, a magnetorotationally-driven explosion is analyzed: Abdikamalov et al. 2014 \cite{Abdikamalov_2014} (Abd+14) AbdA4O01.0.

We study signals modeled as ringdowns (RDs), representing the final stages of BBH mergers. RDs are also expected from the excitation of fundamental modes in neutron stars~\cite{Benhar:2004xg}, but their typical frequencies are around 2-3\,kHz which is beyond our search range. The RD model follows c.f. Eqn.~(3.6) in Ref.~\cite{Abadie:2012rq}:
\begin{equation}
\begin{split}
h_+(t) & \propto \exp{(-t/\tau)}\sin{(2\pi f_0 t)}\\
h_\times(t) & \propto \exp{(-t/\tau)}\cos{(2\pi f_0 t)},
\end{split}
\label{eq:ringdown}
\end{equation}
where $f_0$ is the central frequency, and $\tau$ is the decay time. 
We use $f_0$ to be: 70\,Hz, 235\,Hz, and 849\,Hz, and $\tau$ is chosen to generate signals with half, one, and two cycles.

\begin{figure*}[hbt]
    \includegraphics[width=0.49\linewidth]{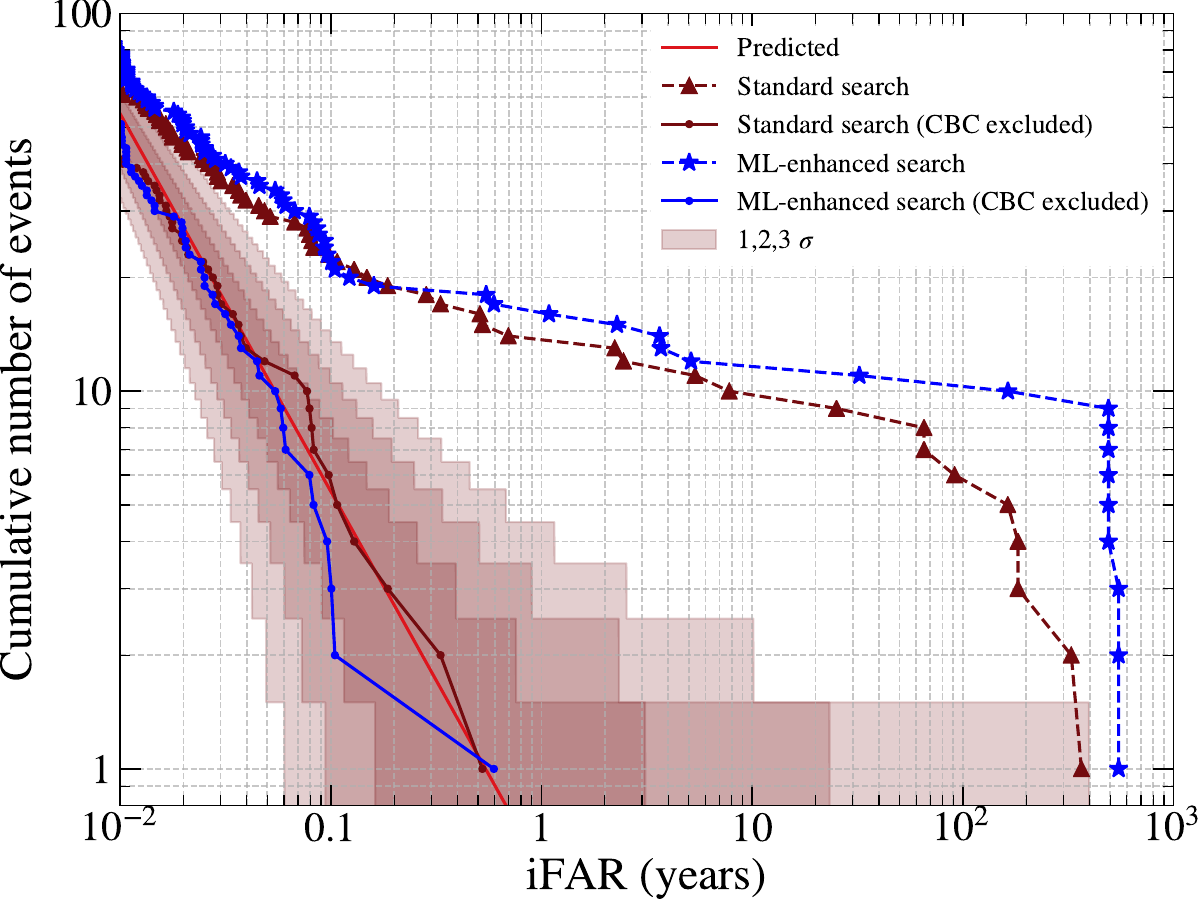}
    \includegraphics[width=0.49\linewidth]{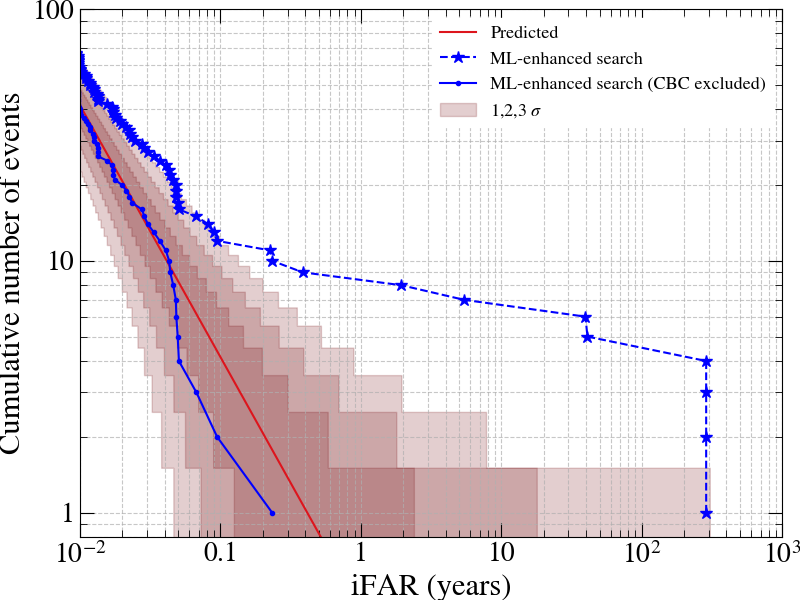}
\caption{Cumulative number of events versus iFAR found by the standard (brown dashed line) and ML-enhanced searches (blue dashed line) in O3. The red solid line shows the expected mean value of the background and the shaded regions are the 1$\sigma$, 2$\sigma$ and 3$\sigma$ Poisson uncertainty intervals. \textit{Left}: results for the HL network. At iFAR~$\geq 1$ year, the ML-enhanced search detected 16 CBC events compared to 14 events for the standard search. \textit{Right}: results for the HLV network, the standard search was not performed, while 10 events are detected with iFAR~$\geq 1$ year. 
In both panels, the loudest events' significance saturates due to the limited amount of background available for testing (1/2 was already used for training), i.e. roughly 500 (300) years for the HL (HLV) network. 
After removing the known CBC events (continuous lines), the ML-enhanced search reports a null result for both networks.}
\label{fig:search}
\end{figure*}

Other potential burst sources are cosmic strings (CS). CSs are one-dimensional topological defects possibly formed after a spontaneous symmetry phase transition in the early Universe and are usually described in grand unified theories. CS cusps and kinks propagating on string loops are expected to generate GW bursts. The CS models allow generating templates and matched filtering was used in the recent LVK search for CSs~\cite{Abbott:2021ksc}. In our analysis, we adopt GW waveforms from CS cusps~\cite{Damour:2004kw}, that has been used in the search for CSs with O1 LIGO data~\cite{LIGOScientific:2017ikf} and are characterized by the amplitude, low-frequency cut-off of 1\,Hz and high-frequency cut-off $f_\mathrm{cutoff}$ of 50\,Hz, 150\,Hz, 500\,Hz and 1500\,Hz. \footnote{The fraction of the CSs' injected amplitudes that fall inside the analyzed frequency band is only 10\%.} 
For the first time, we report the cWB sensitivity on these sources, both for the standard and the ML-enhanced search. 
These results cannot be directly compared to the Ref.~\cite{Abbott:2021ksc} where the detection efficiencies are a function of signal amplitude, not $h_\mathrm{rss}$.

The first set of binaries is BBH systems with quasicircular orbits. The waveforms are calculated using effective-one-body SEOBNRv4 model~\cite{Bohe:2016gbl} that includes only dominant $(l,m)=(2,2)$ mode. The component masses span from 5 $M_\odot$ to 100 $M_\odot$, following the power law + peak~\cite{Talbot:2018cva} mass function. The mass ratio ranges between 1/4 to 1.

The second set of binaries is intermediate-mass black hole systems. Similarly to~\cite{LIGOScientific:2019ysc}, we use 46 IMBH binaries with a total mass between 120~$M_\odot$ and 800~$M_\odot$, and a mass ratio from 1 to 1/10. The systems are precessing, with aligned and anti-aligned spins. These waveforms are  derived from Numerical Relativity simulations (~\cite{Mroue:2013xna, Healy:2017psd, Jani:2016wkt}). Each IMBH binary is uniformly distributed in sky location, inclination angle, and within a co-moving volume, optimized on the signal strength:  
for each signal waveform, we fix 
the maximum redshift by calculating conservative upper bounds on the optimal three-detector network signal-to-noise ratio (SNR) to avoid  generating injections well outside any possible detection range.

The third set contains binaries with eccentric orbits. The eBBH systems could be formed through gravitational capture in close encounters, which causes small orbital separation between the component black holes and high initial orbital eccentricity. We consider 35 mass bins which range between 100 $M_\odot$ and 250 $M_\odot$, mass ratio 1, and eccentricities between 0.19 - 0.96. The simulated eBBH signals used in this analysis are obtained from Numerical Relativity simulations from~\cite{Healy:2017psd}.

\section{Results}
\label{sec:results}

This Section discusses the O3 results of the ML-enhanced cWB search and the sensitivity of the two-fold and three-fold detector networks on the broadest set of signal morphologies performed to date.
We compare our results with the standard generic burst search performed by the LVK collaboration~\cite{O3allskyshort}. More specifically, we used the same standard cWB methods, including the same three search bins to rank separately volumes of signal parameter space dominated by very different false alarm distributions.
Due to the different noise floors and noise transients during O3, the analysis performed by LVK generic burst search considered O3a and O3b separately. In this work, we train two separate XGBoost models for O3a and O3b, respectively. For brevity, all results are reported for the full O3.

\subsection{GW detections}
\label{sec:zero-lag}

All detected GW transients are already established CBC emissions~\cite{GWTC3}, and no evidence is found for other source classes.

\begin{figure*}[hbt]
	\includegraphics[width=1.0\linewidth]{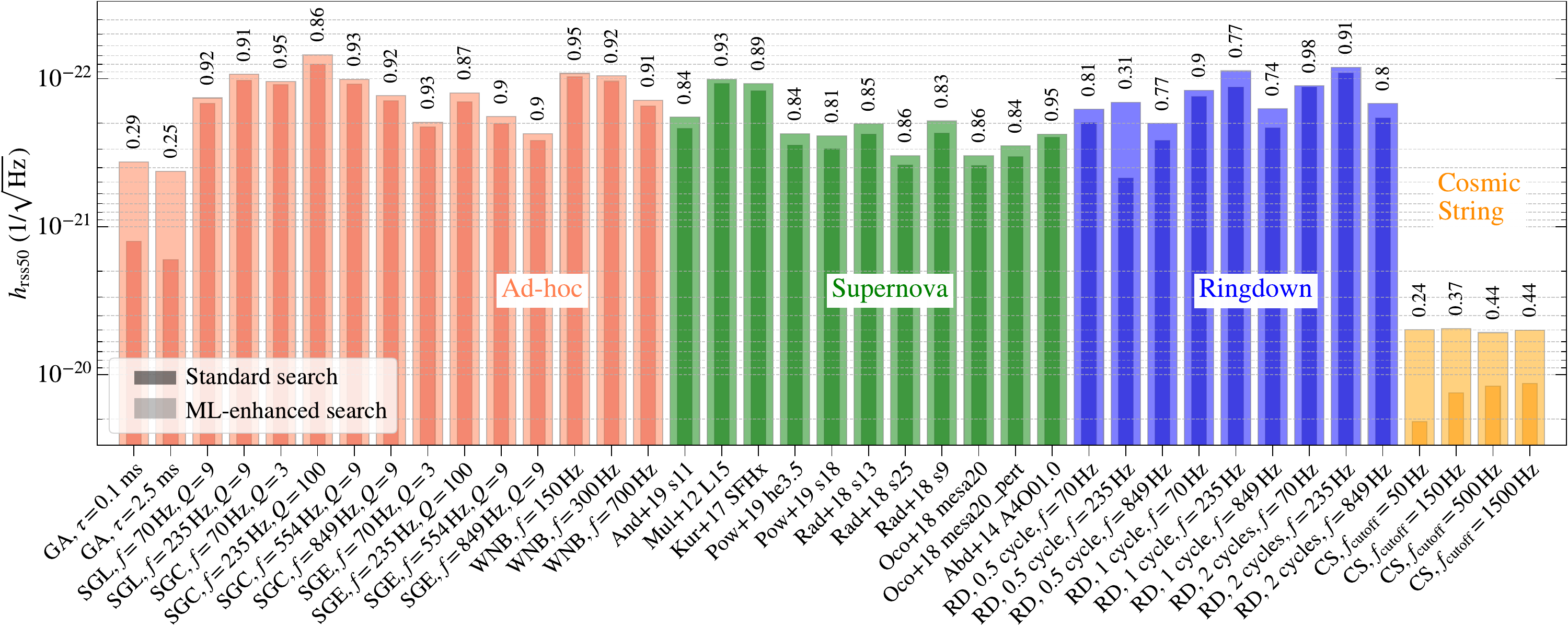}
	\caption{Resulting $h_{\mathrm{rss50}}$ achieved with cWB with standard post-production veto procedure (darker colors) and with  ML-enhanced cWB (lighter colors) for the HL network on full O3 and at iFAR$\geq 100 $ years. The waveforms reported are a subset of those listed in Table~\ref{tab:hrss}: ad-hoc signals ordered according to central frequency (red), core-collapse supernovae (green), ringdown waveforms (blue), and cosmic strings (yellow). The values on the top show the reduction factor on $h_{\mathrm{rss50}}$ with respect to the standard search; $h_{\mathrm{rss50}}$ ordinate scale decreases going upwards.}
	\label{fig:hrss}
\end{figure*}

Figure~\ref{fig:search} presents our search results compared to Ref.~\cite{O3allskyshort} for the HL detector network. At iFAR larger than one year, the present analysis detects more known CBC sources (16) compared to the standard (14), and the significance of 12 out of the 14 common detections is increased. The two CBC detections, which were previously missed by the standard search, are GW190602 with iFAR = 3.7 years, SNR = 11.4 and 48 solar masses, and GW191230 with iFAR = 1.08 years, SNR = 10.4, 36 solar masses. In addition, the ML-enhanced search identified 14 known CBC sources as subthreshold events (iFAR smaller than one year) compared to 7 in the standard analysis. After excluding the known CBC sources, the remaining on-source events are compatible with the estimated background, and the loudest event  shows an iFAR of 0.59 years, comparable to the net observing time (62\% false alarm probability).

This work reports for the first time the O3 detections by a search for GW bursts using the three-detector network, HLV, see right panel of Figure~\ref{fig:search}.
The HLV search detected 8 CBC sources with iFAR larger than 1 per year, with comparable or lower significance with respect to the HL detections. After excluding the known CBC sources, the remaining on-source events are consistent with the estimated background.

\subsection{Sensitivity of two-detector network HL}
\label{sec:HL}
Sensitivity studies show that the ML-enhanced search is systematically improving performances across the tested signal models, therefore preserving the model-agnostic character of the search. Here we focus on the sensitivity achieved on the LIGO Hanford-Livingston network.

\textbf{Amplitude sensitivity}. 
Table~\ref{tab:hrss} and Figure~\ref{fig:hrss} report the comparison between the standard and ML-enhanced searches in terms of typically detectable amplitude, $h_\mathrm{rss50}$ at iFAR$\geq 100$ years, for ad-hoc and astrophysical waveforms. The XGBoost post-production improves the detection efficiency for all the 53 signal morphologies considered.
The greatest improvement is achieved for the waveforms with a few cycles similar to low-Q noise transients (see Sec.~\ref{sec:method}), GAs and CSs. The reduction of $h_\mathrm{rss50}$ with respect to the standard search, $h_\mathrm{rss50,XGB}/h_\mathrm{rss50,STD}$ is around $0.20-0.33$ for GAs and $0.24-0.44$ for CSs. The linear SGs show an improvement of $0.90-0.94$, circular SGs $0.84-0.95$, elliptical SGs $0.87-0.93$, and WNBs $0.91-0.95$. Typically the detectable amplitude improves between 0.30 and 0.98 for ringdowns and  $0.81-0.94$  for CCSN waveforms.
Overall, the ML-enhanced search achieves more homogeneous values of $h_\mathrm{rss50}$, having decreased the gap of sensitivity between the least-performing and best-performing waveform families. In particular, taking into account that for the CSs only 10\% of the injected $h_\mathrm{rss}$ falls inside the analyzed frequency band, amplitude sensitivities to CSs are now comparable to the GA cases.

\begin{figure}[h!]
	\includegraphics[width=0.95\columnwidth]{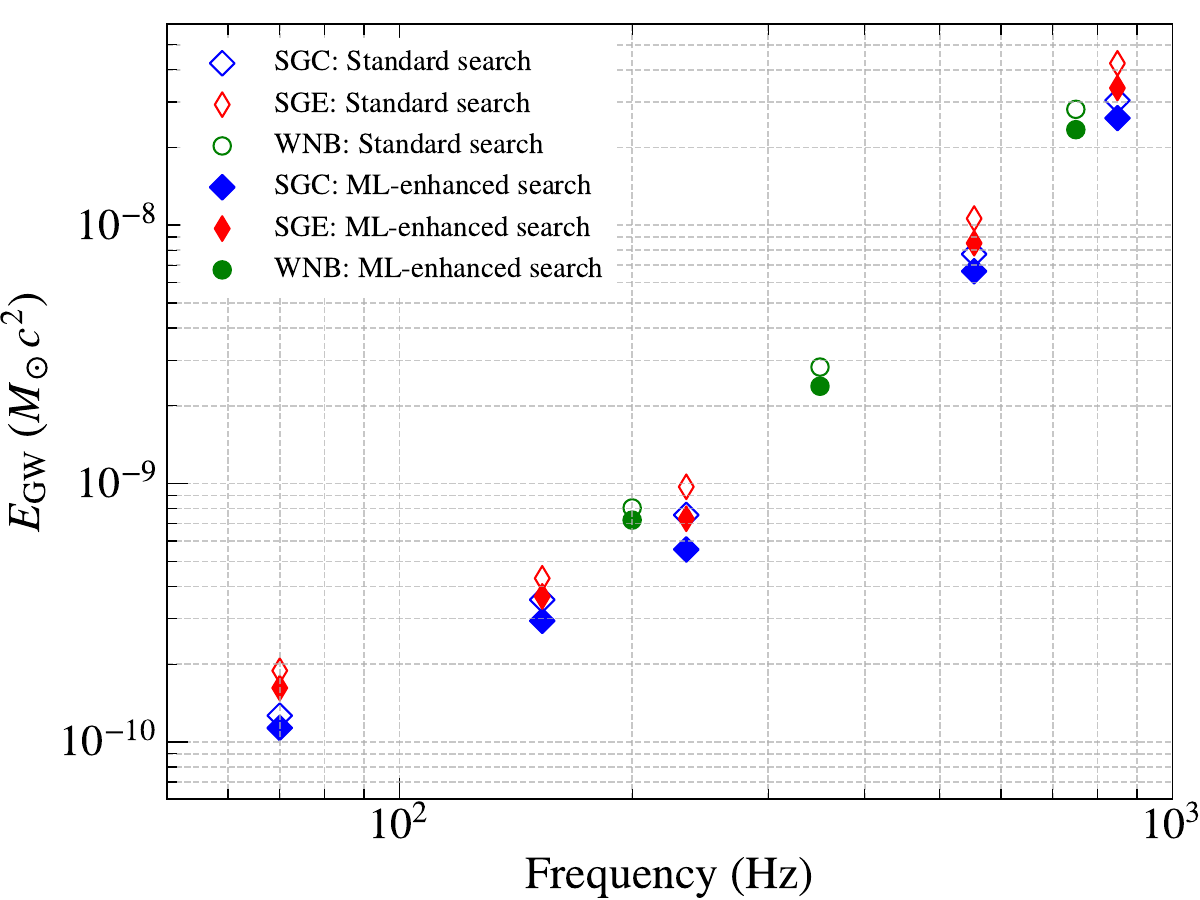}
	\caption{
Radiated energy in GWs at 50\% detection efficiency and iFAR$\geq$ 100~years for a source distance of 10\,kpc. The ML-enhanced cWB improves the constraints across the frequency spectrum for all tested morphologies.}
	\label{fig:egw}
\end{figure}

Figure~\ref{fig:egw} reports the typical detectable energy radiated in GWs for ad-hoc waveforms, evaluated according to the Eqn.~\eqref{eq:eng} by assuming a source distance of 10\,kpc, 50\% detection efficiency and iFAR$\geq$100~years. 
Due to the null detection results for non-CBC GW transients, this can be interpreted as constraints on the product of luminosity distance and amplitude for burst sources. The ML-enhanced cWB improves significantly the constraints across the frequency spectrum for all tested morphologies. 

\begin{figure}[hbt]
	\includegraphics[width=0.95\columnwidth]{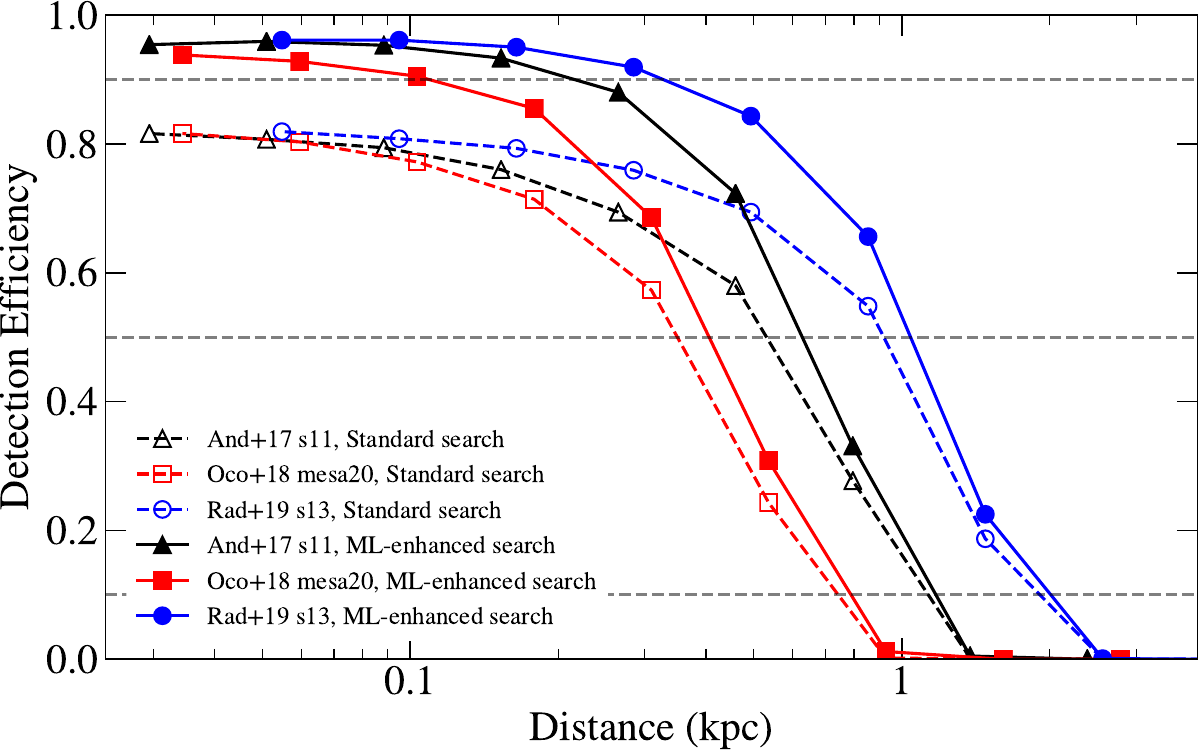}
	\caption{Detection efficiency vs. distance for sample supernova waveforms, for HL network at iFAR$\geq$ 100~years. The ML-enhanced search improves the detection distance at $50\%$ detection efficiency; the probability of detections at a closer distance increases significantly.}
	\label{fig:eff}
\end{figure}

\textbf{Detection range}. As mentioned in Sec.\ref{sec:metric}, in the case of CCSNe models the detection efficiency can be reported as a function of luminosity distance according to Eqn.~\eqref{eqn:dist}. Figure~\ref{fig:eff} shows examples for three CCSN models. The distances at 50\% detection efficiencies are typically around 1\,kpc. Similarly to the improvements in the detectable $h_\mathrm{rss50}$, ML-enhanced cWB allows for an average increase in the detection distance. Moreover, the enhanced algorithm significantly increases the number of detected signals at closer distances.

\begin{figure*}[hbt]
	\includegraphics[width=1\textwidth]
    {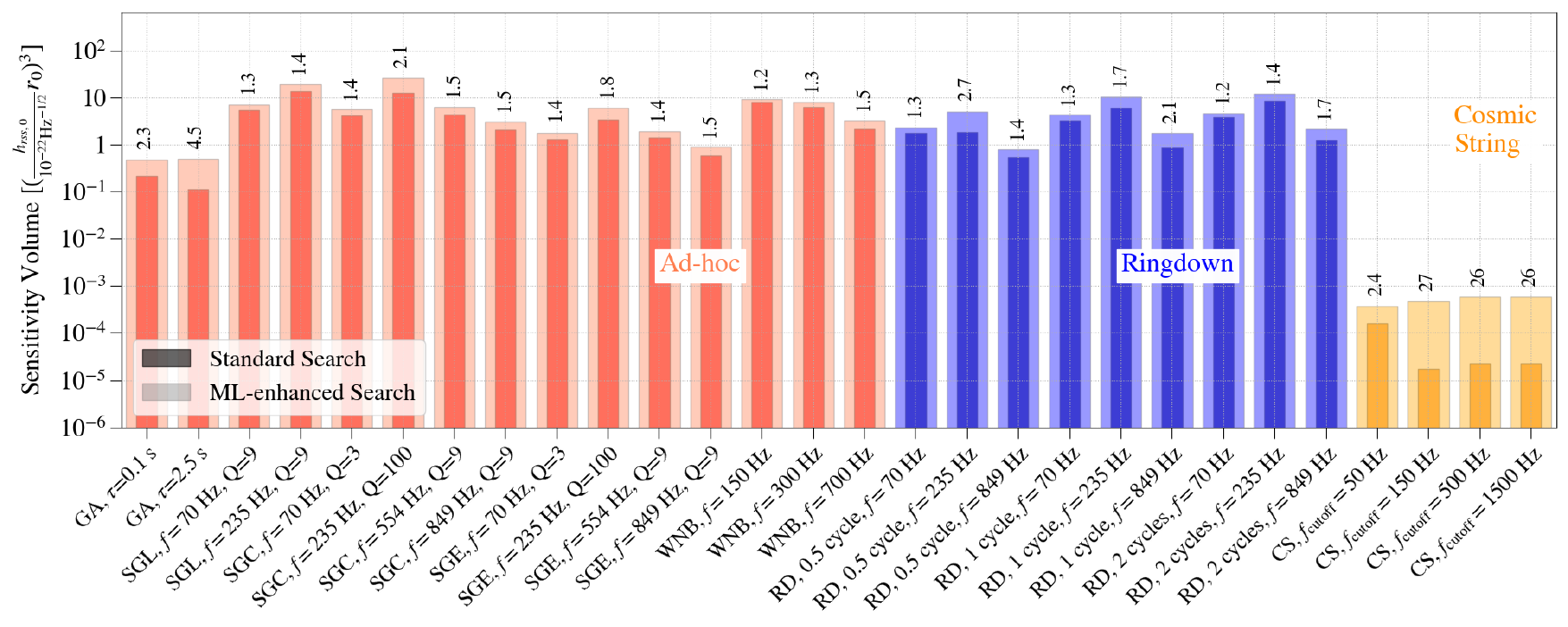}
	\caption{Sensitivity volume obtained with cWB standard post-production veto procedure (darker colors) and with ML-enhanced cWB (lighter colors) for HL network on full O3 data, at iFAR$\geq$ 100 years. The ordinate reports relative sensitive volumes normalized by $4\,\pi\,r_{0}^3$, where $r_{0}$ is the distance at which the source emits the reference amplitude $h_{rss}=10^{-22}Hz^{-1/2}$. We use this standard siren value across all reported signal models to highlight dependencies on signal morphology. 
	The values on the top show the gain in the space volume $\mathcal{V}_\mathrm{XGB}/\mathcal{V}_\mathrm{STD}$. From left to right, the waveforms reported are ad-hoc signals ordered according to frequency (red), 
    ringdown waveforms (blue) and cosmic strings  (yellow).}
	\label{fig:O3_GainVolume}
\end{figure*}

\textbf{Sensitivity volume}. Figure~\ref{fig:O3_GainVolume} presents a volumetric benchmark for ad-hoc and astrophysical waveforms, based on Eqn.~\eqref{eq:vol}. This benchmark is a proxy for the expected detection rate in case of sources uniformly distributed in volume and with a detection range within small cosmological redshifts. 
The sensitivity volume is mainly determined by the detection efficiency $\epsilon (h_\mathrm{rss})$ in the lower $h_\mathrm{rss}$ range, typically well below $h_\mathrm{rss50}$. This benchmark is estimated using Eqn.\ref{eq:vol} by fitting $\epsilon(h_\mathrm{rss})$ over a measured grid of $h_\mathrm{rss}$ values. In a few cases, namely the GA waveforms and the CS $f_\mathrm{cutoff}=50\,\mathrm{Hz}$, the fits fail and Eqn.\ref{eq:vol} is estimated directly from the data points. Results in Fig.~\ref{fig:O3_GainVolume} assume a standard siren model with equal strength across all tested signal models to highlight dependencies on the signal morphology. We do not report sensitivity volumes for CCSNs since each model requires a specific emission strength and current detection ranges are not compatible with the assumption of uniform distribution in volume.

The sensitivity volume shows a systematic improvement with respect to the standard search, with the volume ratio $ \mathcal{V}_{XGB}/ \mathcal{V}_{STD}$ ranging between $1.2 - 1.5$ for WNBs, $1.3 - 2.1$ for SGs and $1.2 - 2.7$ for RDs. Here too, the most substantial improvements are achieved for GAs, $2.3 - 4.5$, and CSs, $2.4 - 27$, confirming that the ML-enhanced search greatly improves its discriminating power against blip glitches and allows to achieve more homogeneous performances across tested signals. The depletion of CSs' results compared to other signals is mostly due to the fact that only a fraction of their injected amplitudes (10\%) falls inside the analyzed frequency band.
 


\begin{table*}[!htb]
\centering
\scriptsize

\begin{tabular}{ c @{\hskip 0.2in} rcl @{\hskip 0.2in} lcl @{\hskip 0.2in} rcl @{\hskip 0.2in} lcl }

\hline
\hline


\textbf{} & \multicolumn{6}{c}{\textbf{HL network}} & \multicolumn{6}{c}{\textbf{HLV network}} \\
\textbf{Morphology} 
& \multicolumn{3}{c}{$h_\mathrm{rss50}$ $[1/\sqrt{\mathrm{Hz}}]$} 
& \multicolumn{3}{c}{$\mathcal{V}$ [$(\frac{h_\mathrm{rss,0}}{1e-22}r_0)^3$]} 
& \multicolumn{3}{c}{$h_\mathrm{rss50}$ $[1/\sqrt{\mathrm{Hz}}]$}
& \multicolumn{3}{c}{$\mathcal{V}$ [$(\frac{h_\mathrm{rss,0}}{1e-22}r_0)^3$]} \\
\textbf{} 
& \textbf{STD} &/& \textbf{XGB}
& \textbf{STD} &/& \textbf{XGB}
& \textbf{STD} &/& \textbf{XGB}
& \textbf{STD} &/& \textbf{XGB} \\

\hline

\textbf{Gaussian pulse}\\
$\tau=4.0$\,ms & 27.0 &/& 5.5 & 9.6e--2 $^{\dagger}$ &/& 2.5e--1 $^{\dagger}$ & 94.5 &/& 13.8 & 7.2e--2 $^{\dagger}$ &/& 8.2e--2 $^{\dagger}$ \\
$\tau=2.5$\,ms & 16.7 &/& 4.2 & 1.1e--1 $^{\dagger}$ &/& 5.1e--1 $^{\dagger}$ & 31.8 &/& 10.9 & 1.7e--1 $^{\dagger}$ &/& 1.2e--1 $^{\dagger}$ \\
$\tau=1.0$\,ms & 11.1 &/& 3.7 & 1.1e--1 $^{\dagger}$ &/& 1.5 $^{\dagger}$ & 13.9 &/& 8.4 & 9.7e--2 $^{\dagger}$ &/& 3.5e--1 $^{\dagger}$\\
$\tau=0.1$\,ms & 12.6 &/& 3.6 & 2.1e--1 $^{\dagger}$ &/& 4.9e--1 $^{\dagger}$ & 17.5 &/& 11.7 & 1.7e--1 $^{\dagger}$ &/& 9.9e--2 $^{\dagger}$\\
\textbf{Sine-Gaussian linear}\\
$f=70\,\mathrm{Hz},  Q=9$ & 1.5 &/& 1.4 & 5.5 &/& 7.1 & 1.4 &/& 1.4 & 5.0 &/& 5.8 \\
$f=100\,\mathrm{Hz}, Q=9$ & 1.2 &/& 1.1 & 9.8 &/& 1.2e+1 & 1.2 &/& 1.2 & 8.2 &/& 9.0 \\
$f=235\,\mathrm{Hz}, Q=9$ & 1.0 &/& 0.9 & 1.4e+1 &/& 1.9e+1 & 1.1 &/& 1.0 & 1.1e+1 &/& 1.4e+1 \\
$f=361\,\mathrm{Hz}, Q=9$ & 1.2 &/& 1.1 & 1.0e+1 &/& 1.5e+1 & 1.3 &/& 1.2 & 7.9 &/& 1.0e+1 \\
\textbf{Sine-Gaussian circular}\\
$f=70\,\mathrm{Hz},  Q=3$ & 1.1 &/& 1.0 & 4.1 &/& 5.6 & 1.2 &/& 1.2 & 3.4 &/& 4.3 \\
$f=70\,\mathrm{Hz},  Q=100$ & 1.1 &/& 1.0 & 4.5 &/& 6.6 & 1.1 &/& 1.0 & 4.4 &/& 5.6 \\
$f=153\,\mathrm{Hz}, Q=9$ & 0.8 &/& 0.8 & 1.1e+1 &/& 1.6e+1 & 0.9 &/& 0.9 & 9.2 &/& 1.1e+1 \\
$f=235\,\mathrm{Hz}, Q=3$ & 0.9 &/& 0.8 & 9.9 &/& 1.4e+1 & 0.9 &/& 0.9 & 8.1 &/& 8.8 \\
$f=235\,\mathrm{Hz}, Q=100$ & 0.8 &/& 0.7 & 1.3e+1 &/& 2.6e+1 & 0.8 &/& 0.7 & 1.1e+1 &/& 1.6e+1 \\
$f=554\,\mathrm{Hz}, Q=9$ & 1.1 &/& 1.0 & 4.3 &/& 6.2 & 1.2 &/& 1.2 & 3.5 &/& 4.2 \\
$f=849\,\mathrm{Hz}, Q=3$ & 1.6 &/& 1.5 & 1.5 &/& 1.9 & 1.8 &/& 1.7 & 1.1 &/& 1.2 \\
$f=849\,\mathrm{Hz}, Q=9$ & 1.4 &/& 1.3 & 2.1 &/& 3.1 & 1.6 &/& 1.5 & 1.6 &/& 1.9 \\
$f=849\,\mathrm{Hz}, Q=100$ & 1.4 &/& 1.2 & 2.2 &/& 3.8 & 1.5 &/& 1.4 & 1.9 &/& 2.7 \\
\textbf{Sine-Gaussian elliptical}\\
$f=70\,\mathrm{Hz},  Q=3$ & 2.1 &/& 2.0 & 1.3 &/& 1.8 & 2.3 &/& 2.2 & 1.0 &/& 1.2 \\
$f=70\,\mathrm{Hz},  Q=100$ & 2.0 &/& 1.8 & 1.5 &/& 2.0 & 1.9 &/& 1.7 & 1.4 &/& 1.3 \\
$f=153\,\mathrm{Hz}, Q=9$ & 1.5 &/& 1.4 & 3.4 &/& 4.8 & 1.6 &/& 1.5 & 2.9 &/& 3.3 \\
$f=235\,\mathrm{Hz}, Q=3$ & 1.6 &/& 1.5 & 2.6 &/& 3.6 & 1.7 &/& 1.8 & 2.5 &/& 2.7 \\
$f=235\,\mathrm{Hz}, Q=100$ & 1.4 &/& 1.2 & 3.4 &/& 6.0 & 1.4 &/& 1.3 & 3.4 &/& 4.5 \\
$f=554\,\mathrm{Hz}, Q=9$ & 2.0 &/& 1.8 & 1.4 &/& 1.9 & 2.3 &/& 2.1 & 1.0 &/& 1.3 \\
$f=849\,\mathrm{Hz}, Q=3$ & 2.9 &/& 2.7 & 4.5e--1 &/& 5.5e--1 & 3.5 &/& 3.2 & 3.4e--1 &/& 4.0e--1 \\
$f=849\,\mathrm{Hz}, Q=9$ & 2.6 &/& 2.4 & 5.9e--1 &/& 8.8e--1 & 3.0 &/& 2.8 & 4.8e--1 &/& 5.5e--1 \\
$f=849\,\mathrm{Hz}, Q=100$ & 2.6 &/& 2.3 & 6.3e--1 &/& 1.1 & 2.8 &/& 2.6 & 5.6e--1 &/& 7.1e--1 \\
\textbf{White Noise Burst}\\
$f=150\,\mathrm{Hz}$ & 1.0 &/& 0.9 & 7.8 &/& 9.2 & 1.1 &/& 1.0 & 6.1 &/& 6.6 \\
$f=300\,\mathrm{Hz}$ & 1.0 &/& 1.0 & 6.2 &/& 8.0 & 1.2 &/& 1.1 & 4.9 &/& 5.8 \\
$f=700\,\mathrm{Hz}$ & 1.5 &/& 1.4 & 2.2 &/& 3.3 & 1.8 &/& 1.5 & 1.6 &/& 2.0 \\
\textbf{Supernova}\\
Mul+12 L15 & 1.1 &/& 1.0 & \multicolumn{3}{c}{-} & 1.2 &/& 1.1 & \multicolumn{3}{c}{-} \\
Abd+14 A4O01.0 & 2.5 &/& 2.4 & \multicolumn{3}{c}{-} & 3.0 &/& 13.1 & \multicolumn{3}{c}{-} \\
Kur+17 SFHx & 1.2 &/& 1.1 & \multicolumn{3}{c}{-} & 1.4 &/& 1.2 & \multicolumn{3}{c}{-} \\
Rad+18 s9 & 2.3 &/& 1.9 & \multicolumn{3}{c}{-} & 3.3 &/& 2.3 & \multicolumn{3}{c}{-} \\
Rad+18 s13 & 2.4 &/& 2.0 & \multicolumn{3}{c}{-} & 3.1 &/& 2.2 & \multicolumn{3}{c}{-} \\
And+19 s11 & 2.2 &/& 1.8 & \multicolumn{3}{c}{-} & 2.9 &/& 2.2 & \multicolumn{3}{c}{-} \\
Oco+18 mesa20$\_$pert & 3.4 &/& 2.8 & \multicolumn{3}{c}{-} & 4.9 &/& 3.5 & \multicolumn{3}{c}{-} \\
Oco+18 mesa20 & 3.9 &/& 3.3 & \multicolumn{3}{c}{-} & 5.5 &/& 4.7 & \multicolumn{3}{c}{-} \\
Pow+19 he3.5 & 2.8 &/& 2.4 & \multicolumn{3}{c}{-} & 3.9 &/& 2.6 & \multicolumn{3}{c}{-} \\
Rad+18 s25 & 3.9 &/& 3.3 & \multicolumn{3}{c}{-} & 5.1 &/& 3.7 & \multicolumn{3}{c}{-} \\
Pow+19 s18 & 3.0 &/& 2.4 & \multicolumn{3}{c}{-} & 4.2 &/& 2.7 & \multicolumn{3}{c}{-} \\
\textbf{Ringdown}\\
0.5 cycle, $f=70\,\mathrm{Hz}$ & 2.0 &/& 1.6 & 1.8 &/& 2.4 & 3.5 &/& 7.6 & 4.1e--1 &/& 3.1e--1 \\
1 cycle, $f=70\,\mathrm{Hz}$ & 1.3 &/& 1.2 & 3.2 &/& 4.2 & 2.6 &/& 3.8 & 6.7e--1 &/& 7.8e--1 \\
2 cycles, $f=70\,\mathrm{Hz}$ & 1.1 &/& 1.1 & 3.8 &/& 4.7 & 2.3 &/& 1.8 & 9.9e--1 &/& 1.3 \\
0.5 cycle, $f=235\,\mathrm{Hz}$ & 4.7 &/& 1.4 & 1.9 &/& 5.1 & 7.3 &/& 4.0 & 1.1 &/& 1.8 \\
1 cycle, $f=235\,\mathrm{Hz}$ & 1.1 &/& 0.9 & 6.0 &/& 1.0e+1 & 1.3 &/& 1.1 & 5.2 &/& 6.9 \\
2 cycles, $f=235\,\mathrm{Hz}$ & 0.9 &/& 0.8 & 8.6 &/& 1.2e+1 & 1.0 &/& 0.9 & 7.5 &/& 8.9 \\
0.5 cycle, $f=849\,\mathrm{Hz}$ & 2.6 &/& 2.0 & 5.5e--1 &/& 7.8e--1 & 2.3 &/& 3.1 & 1.4 &/& 1.3 \\
1 cycle, $f=849\,\mathrm{Hz}$ & 2.2 &/& 1.6 & 8.5e--1 &/& 1.8 & 1.4 &/& 1.3 & 2.6 &/& 3.1 \\
2 cycles, $f=849\,\mathrm{Hz}$ & 1.8 &/& 1.5 & 1.2 &/& 2.1 & 1.2 &/& 1.2 & 3.3 &/& 3.8 \\
\textbf{Cosmic String}\\
$f_\mathrm{cutoff}=50\,\mathrm{Hz}$ & 208.3 &/& 49.8 & 1.6e--4 $^{\dagger}$ &/& 3.8e--4 $^{\dagger}$ & 336.4 &/& 246.7 & 1.8e--4 $^{\dagger}$ &/& 9.2e--5 $^{\dagger}$ \\
$f_\mathrm{cutoff}=150\,\mathrm{Hz}$ & 133.5 &/& 48.8 & 1.7e--5 &/& 4.7e--4 & 180.2 &/& 117.9 & 9.0e--6 &/& 1.3e--4 \\
$f_\mathrm{cutoff}=500\,\mathrm{Hz}$ & 119.6 &/& 52.0 & 2.3e--5 &/& 6.0e--4 & 155.6 &/& 114.0 & 1.1e--5 &/& 1.7e--4 \\
$f_\mathrm{cutoff}=1500\,\mathrm{Hz}$ & 114.6 &/& 50.1 & 2.3e--5 &/& 6.0e--4 & 148.4 &/& 106.5 & 1.1e--5 &/& 1.7e--4 \\

\hline
\hline
\end{tabular}
\caption{The $h_\text{rss}$ values (in units of $10^{-22}$ $\text{Hz}^{-1/2}$) for which 50\% detection efficiency and sensitive volumes $\mathcal{V}$ (in units of [$(\frac{h_\mathrm{rss,0}}{1e-22}r_0)^3$]) is achieved with an iFAR of 100 years for each of the injected signal morphologies. We combine O3a and O3b data, STD and XGB stands for Standard search and ML-enhanced search, respectively. For the signal morphology marked with $^{\dagger}$, the $h_{\mathrm{rss}50}$ value for the standard cWB search was estimated directly using the data points, not from the detection efficiency fit of the data (for more details see Section \ref{sec:HL}, sensitivity volume).}
\label{tab:hrss}
\end{table*}

\begin{figure}[]
	\includegraphics[width=0.95\columnwidth]{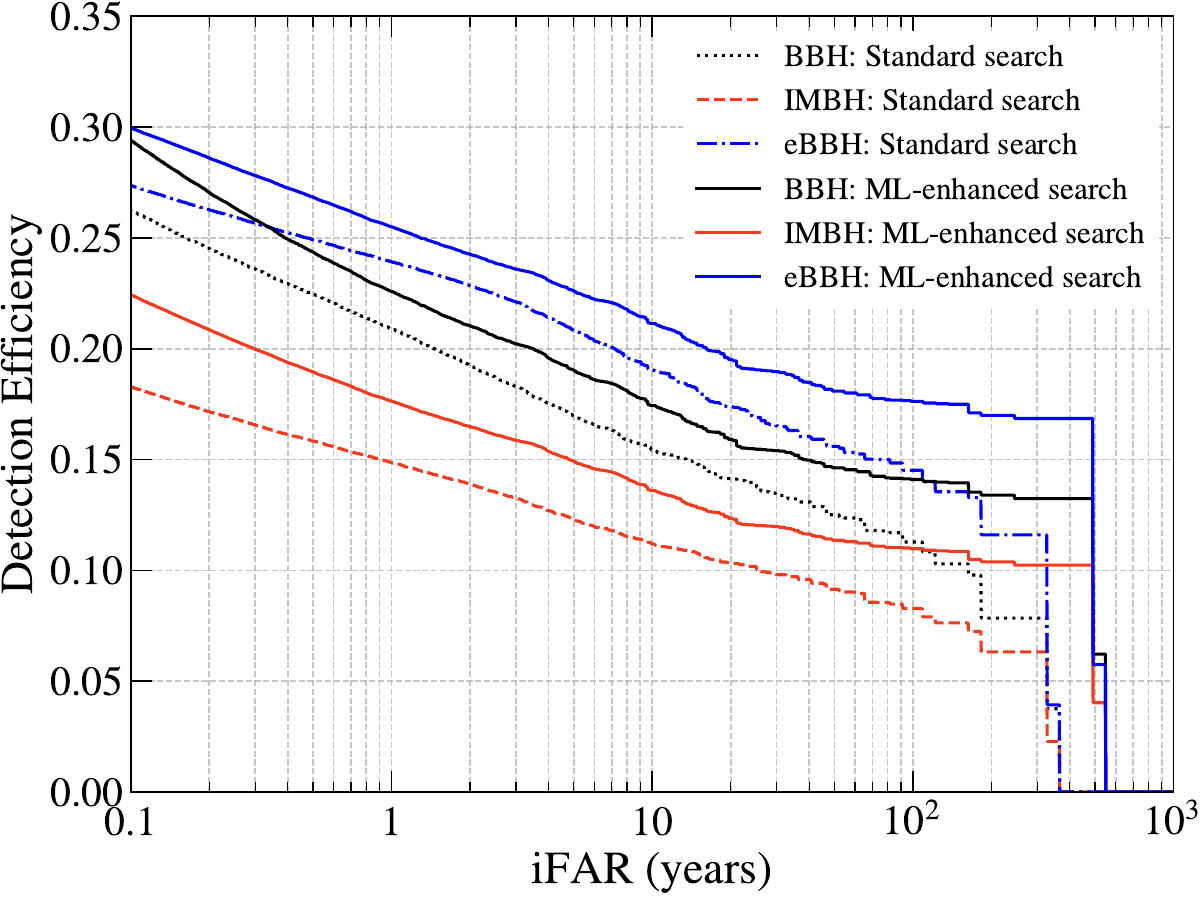}
	\caption{Detection efficiency versus iFAR for BBH 
	(black), IMBH (red) and eBBH (blue). ML-enhanced cWB (solid lines) shows an increase in detection efficiency with respect to the standard search (dashed lines).
	}
	\label{fig:roc}
\end{figure}

\textbf{Receiver operating characteristics curves for compact binaries}. 
We extend the robustness test to show that the ML-enhanced cWB for generic transients is also providing a better sensitivity to CBC sources than the standard method for bursts. This test is performed on three CBC simulation sets, stellar mass BBHs, IMBHBs, and eBBHs, each one spanning different regions in parameter spaces and fiducial volumes as described in Sec.~\ref{subsec:astro}. The receiver operating characteristics curves for each set are shown in Figure~\ref{fig:roc}. Detection efficiencies improve across the entire iFAR threshold range, detecting around $8 - 25\%$ more from our BBH set, $18 - 32\%$ more from our IMBH set, and $8 - 21\%$ more from our eBBH set.\footnote{Our receiver operating characteristic curves cannot be compared across different sets, nor are meant to be representative of a sensitive volume for each set. This is so because each set is built under very different assumptions on rate densities and distributions of intrinsic parameters of the sources.} The results of this test are found to be consistent with the increased number of CBC detections presented in Sec.~\ref{sec:zero-lag}. 


\subsection{Sensitivity of three-detector network HLV}

\begin{figure*}[hbt!]
	\includegraphics[width=1\textwidth]
    {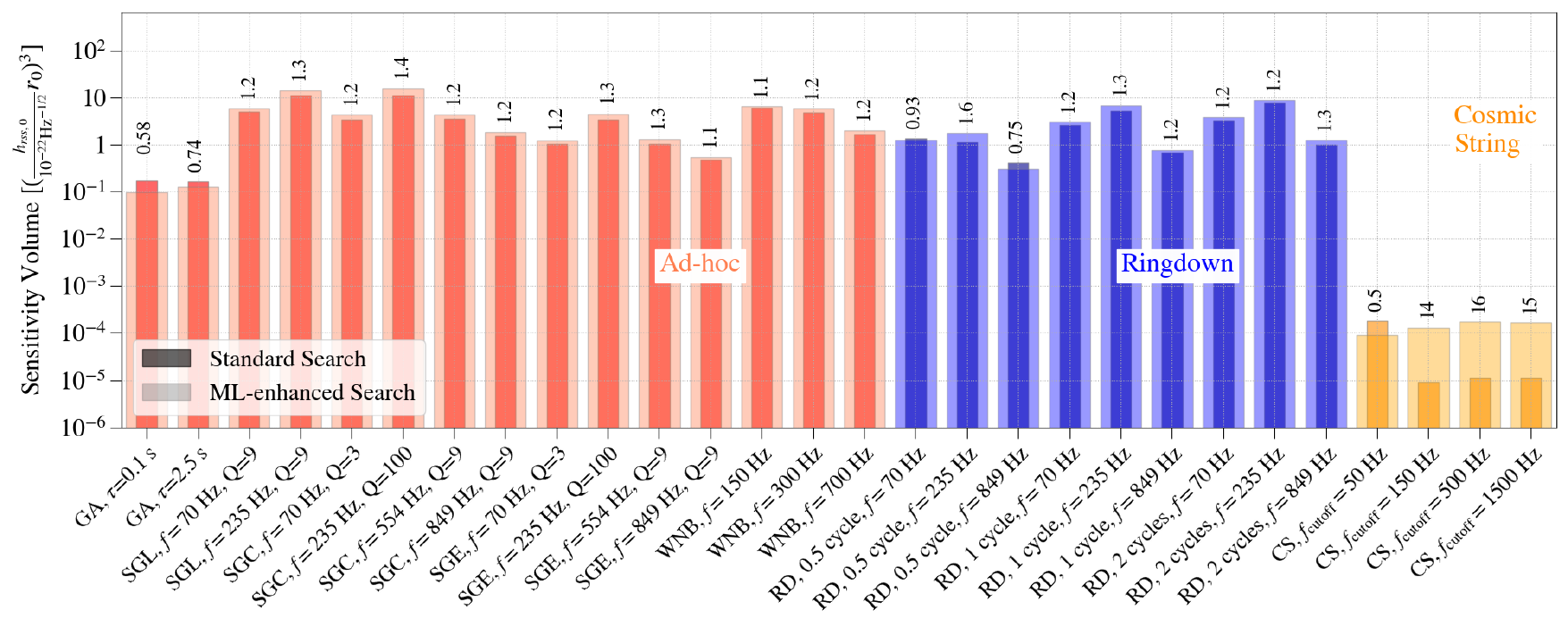}
 \caption{Sensitivity volume obtained with cWB standard post-production veto procedure (darker colors) and with cWB ML-enhanced (lighter colors) for HLV network during full O3 run, at iFAR=100 years. The values above each column show the space volume gain $\mathcal{V}_\mathrm{XGB}/\mathcal{V}_\mathrm{STD}$. From left to right, the waveforms reported are: ad-hoc signals ordered according to frequency (red), 
 ringdown waveforms (blue) and cosmic strings (yellow).}
	\label{fig:O3_GainVolume_HLV}
\end{figure*}

We test the ML-enhanced cWB on HLV data using the same set of simulated signals as for the HL network. For brevity, we report the results only in terms of $h_\mathrm{rss50}$ and sensitivity volume (Table~\ref{tab:hrss}, Figure~\ref{fig:O3_GainVolume_HLV}). Compared to the standard HLV analysis, the ML-enhanced cWB improves the $h_\mathrm{rss50}$ for 48 
out of the 53 considered waveforms~\footnote{While a light degradation of performances for a few waveforms can be considered acceptable when using ML-enhanced cWB, in the case of the Abdikamalov supernova waveform \cite{Abdikamalov_2014}, the $h_\mathrm{rss50}$ increases by a factor of 4. The simulated events 
for such waveform have a time-frequency evolution similar to short-duration transient noise and belong to a Qa-Qp region not entirely covered by our signal training set. This resulted in a poor separation from blip glitches on O3b data in the HLV configuration, see Appendix~\ref{sec:xgb}.}. 
Also, the sensitive volume $\mathcal{V}$ improves for 36 out of the 42 considered waveforms.

Analyzing the HLV results for O3a and O3b separately, we observe that ML-enhanced cWB search demonstrates better performance in O3a. The O3b analysis is more challenging due to a significantly higher rate of short-duration noise transients.
We remark that in order to take full advantage of the addition of Virgo data, the analysis should search for both GW polarization components. In order to preserve the generality of the search, those polarization components cannot be assumed to be correlated. This is not the case for the HL network, where both instruments sense approximately the same GW polarization component from most sky directions. Therefore, the HLV analysis provides a less efficient discrimination of glitches, resulting in a lower detection efficiency of GW signals at a given statistical confidence.



\section{Conclusions}
\label{sec:conclusions}

This paper reports a search for  generic GW burst sources on the data from the third observing run of the Advanced LIGO and Virgo detectors. We use coherent WaveBurst pipeline enhanced with machine learning to discriminate noise from signal events. The ML algorithm is trained using generic signals with stochastic morphology that do not match any known signal model. This procedure preserves the model-independent nature of the search.

The ML enhancement substantially improves the all-sky search for burst sources. We achieved a systematic improvement of sensitivity across the very broad set of tested signal morphologies, including ad-hoc signal classes and waveforms from astrophysical sources (core-collapse supernovae, cosmic strings and ringdown waveforms). 
The most significant improvement in sensitivity compared to the LVK O3 search~\cite{O3allskyshort} is achieved for low-Q factor signals, such as cosmic strings and Gaussian pulses. In fact, the ML-enhanced algorithm is more successful in discriminating the dominant class of noise events such as blip glitches. 
Detection performances are measured in terms of the detectable signal amplitudes, distance ranges and space volumes. In particular, detectable amplitudes improve from a few percent to almost an order of magnitude. 
The ML-enhanced cWB is also more efficient in reconstructing compact binary coalescences with stellar mass and intermediate-mass black holes. These tests demonstrate that the algorithm robustly detects a wide range of short-duration GW transients. 

This search detects more compact binaries than the previous cWB burst search, improving their significance. The results of the three-fold LIGO-Virgo detector network with O3 data are reported here for the first time. After the compact binaries are excised, we report no evidence for new GW transients for both two-fold LIGO and three-fold LIGO-Virgo detector networks.

\begin{acknowledgments}

We gratefully acknowledge the computational resources provided by LIGO-Virgo. This material is based upon work supported by NSF's LIGO Laboratory which is a major facility fully funded by the National Science Foundation. This research has made use of data, software and/or web tools obtained from the Gravitational Wave Open Science Center, a service of LIGO Laboratory, the LIGO Scientific Collaboration and the Virgo Collaboration. The work by S. K. was supported by NSF Grants No. PHY~1806165 and PHY~2110060. I.B. acknowledges the support of the Alfred P. Sloan Foundation and NSF grants PHY-1911796 and PHY-2110060. A.M. acknowledges the support of the European Gravitational Observatory, convention EGO-DIR-63-2018. S.T. is supported by Swiss National Science Foundation (SNSF) Ambizione Grant Number : PZ00P2\_202204. 

\end{acknowledgments}

\appendix

\section{XGBoost Model}
\label{sec:xgb}

The ML-enhanced cWB search was introduced in Ref.~\cite{XGBcwb} and updated for the O3 BBH search in Ref.~\cite{O3XGB}. We use a decision tree-based ensemble learning classifying algorithm called eXtreme-Gradient Boost (XGBoost). The cWB pipeline is utilized to reconstruct interesting events and generate summary statistics. A carefully selected subset of summary statistics is used as the list of input features for the ML model. 

In this subsection, we discuss the necessary updates to the BBH/IMBH ML-enhanced cWB search~\cite{O3XGB}, in order to tune it for the generic burst search:

While the cWB search described in Ref.~\cite{O3XGB} looks for compact binary coalescence signals that do not depend on accurate waveform modeling, the generic searches should be designed for an even wider range of burst morphologies. The definitions of the summary statistics used for the O3 BBH burst search can be found in Ref.~\cite{XGBcwb}. We use the following subset of the summary statistics to train XGBoost: norm ($n_\mathrm{f}$), cross-correlation coefficient ($c_\mathrm{c}$), quality of event reconstruction ($\chi^2$), square of SNR over likelihood ($\mathrm{\textit{SNR}}_\mathrm{i}/L$; where i = 0 for the HL network, and i = \{0, 1\} for the HLV network), effective correlated SNR ($\eta_\mathrm{0}$). 
Additionally, we use shape parameters ($Q_\mathrm{a}$, $Q_\mathrm{p}$) which are developed to identify short-duration low-Q glitches.
We exclude model-dependent summary statistics: central frequency ($f_0$), duration ($\Delta T_\mathrm{s}$), and bandwidth ($\Delta F_\mathrm{s}$) as they are strongly correlated with the signal parameters like total/component masses. We also exclude the chirp parameters (chirp mass $\mathcal{M}$, chirp ellipticity $e_{\mathrm{M}}$) since they depend on the time-frequency evolution of the signal.

We use White-Noise-Burst (WNB) signals to train XGBoost, as WNBs allow probing of different regions in the time-frequency map where we expect to find GW signals. Training XGBoost on WNBs ensures the model independence of the ML-enhanced cWB generic burst search as the training set is completely independent of anticipated GW signal models. We generate two sets of WNB for training:\\ (a) WNBs uniformly distributed in central frequency in the range [24,996] Hz, bandwidth [10,300]\,Hz, and duration is logarithmically distributed between 0.1\,ms and 1\,ms.\\ (b) WNBs logarithmically distributed in central frequency [24,450]\,Hz, bandwidth [10,250]\,Hz and duration [1,50]\,ms.\\ The choice of the WNB signals is largely motivated by looking at the $Q_\mathrm{a}$-$Q_\mathrm{p}$ parameters and ensuring that there are no gaps left in the parameter space where we expect to find GW events. Using WNB signals instead of ad-hoc GW waveform signals helps in not biasing the ML algorithm to specific signal models.

We use 50\% of the accumulated background for each run and generate the above-mentioned WNB signals for training. Testing is done on various GW signal simulations and the remaining background for O3a and O3b, respectively.

We expect to encounter GW signals with very different waveform morphology than BBH/IMBH. 
We start by extending the cap for the $\eta_\mathrm{0}$ statistic to $\eta_\mathrm{0} = 20$ for O3a and $25$ for O3b (which was earlier set to $11$ for BBH/IMBH searches in the O1-O2 reanalysis paper~\cite{XGBcwb}, and to $20$ in the O3 reanalysis paper~\cite{O3XGB}). Since the number of summary statistics used as input features was halved, we change the XGBoost hyper-parameter \texttt{max\_depth} from 13 to 6, and \texttt{min\_child\_weight} was optimized from 10 to 5. All the other XGBoost hyper-parameters are the same as reported in the O3 reanalysis paper for the BBH/IMBH search~\cite{O3XGB}. 
With these changes, short grid searches were performed to determine the optimal value for the user-defined weight options (introduced in Ref.~\cite{XGBcwb}) for O3a ($q=5$, $A=40$) and O3b ($q=6$, $A=80$). The list of final XGBoost hyper-parameter values can be found in Table~\ref{tab:xgb}. The same XGBoost hyper-parameters are used for both the two-fold (HL) and three-fold (HLV) networks. 
\begin{table}[]
    \centering
    \begin{tabular}{cc}
        \hline
        \hline
        XGBoost hyper-parameter & entry \\
        \hline
        \hline
        \texttt{objective} & \texttt{binary:logistic}  \\
        \texttt{tree\_method} & \texttt{hist} \\
        \texttt{grow\_policy} & \texttt{lossguide} \\
        \texttt{n\_estimators} & 20,000$\dagger$ \\
        \texttt{max\_depth} & 6 \\
        \texttt{learning\_rate} & 0.03 \\
        \texttt{min\_child\_weight} & 5.0 \\
        \texttt{colsample\_bytree} & 1.0 \\
        \texttt{subsample} & 0.6 \\
        \texttt{gamma} & 2.0 \\
        \hline
        \hline
    \end{tabular}
    \caption{Entries for XGBoost hyper-parameters. $\dagger$: \texttt{n\_estimators} is optimized using early stopping, where the training stops when the validation score stops improving.}
    \label{tab:xgb}
\end{table}

A further correction is applied to the ML penalty 
factor $P_{\mathrm{XGB}}$  
to reduce the high SNR background outliers. $P_{\mathrm{XGB}}$ is ranking criteria obtained from XGBoost (see Ref.~\cite{XGBcwb}). 
These high SNR background outliers usually correspond to a low value of $Q_\mathrm{a}$ and $Q_\mathrm{p}$ (especially blip glitches), and we can suppress them in the affected parameter space by applying the following correction:
\begin{equation}
\footnotesize
P^{'}_{\mathrm{XGB}} = 
\begin{cases}
    P_{\mathrm{XGB}} - \alpha ( 0.15 + \Delta Q_{a,p}) \hspace{1.1cm} \text{if } \ \Delta Q_{a,p} \leq 0.15\\
    P_{\mathrm{XGB}} \hspace{3.65cm} \text{if } \ \Delta Q_{a,p} > 0.15\\
\end{cases}
\end{equation}
where $\alpha$ is usually equal to 1, except for HLV network on O3b data where $\alpha$=3, due to the higher transient noise rate and $\Delta Q_{a,p} = Q_a (Q_p - 0.6)$ defines the penalization curve in the $Q_\mathrm{a} - Q_\mathrm{p}$ plane. 
More details can be found in Ref.~\cite{XGBcwb}. The factor $W_{\mathrm{XGB}}$ in Eq.~\ref{eq:x} is a monotonic function of $P^{'}_{\mathrm{XGB}}$ to increase the dynamic range.

\newpage
\bibliography{paper.bib}

\end{document}